\documentclass[pra,twocolumn,amsmath,amssymb]{revtex4-1}
\usepackage[T1]{fontenc} % for Rzazewski
\usepackage{bm}
\usepackage{graphicx}
\usepackage[bookmarks=false]{hyperref}
\def\ch#1{{#1}}

\newcommand{\ket}[1]{|#1\rangle}
\newcommand{\braket}[1]{\langle #1 \rangle}

\def\dd{\mathrm{d}}
\def\ee{\mathrm{e}}
\def\ii{\mathrm{i}}
\def\vnabla{\bm{\nabla}}

\def\rot{\vnabla\times}

\def\Tr{\mathrm{Tr}}

\def\diez{\varepsilon_0}
\def\muz{\mu_0}
\def\wc{\omega_{\text{r}}}
\def\wa{\omega_{\text{a}}}

\def\LR{L_{\text{r}}}
\def\CR{C_{\text{r}}}
\def\ZR{Z_{\text{r}}}
\def\Lc{L_{\text{c}}}

\def\LT{L_{\text{t}}}
\def\CT{C_{\text{t}}}
\def\dx{\varDelta x}
\def\EJ{E_{\text{J}}}
\def\CJ{C_{\text{J}}}

\def\fluxq{\Phi_0}
\def\Phiext{\Phi_{\text{ext}}}

\def\LL{\mathcal{L}}
\def\LLblack{\mathcal{L}_{\text{black}}}

\def\HH{\mathcal{H}}
\def\HHblack{\mathcal{H}_{\text{black}}}

\def\oH{\hat{\mathcal{H}}}

\def\oHmin{\hat{\mathcal{H}}_{\text{min}}}
\def\oHdipole{\hat{\mathcal{H}}_{\text{dip}}}
\def\oHatom{\hat{\mathcal{H}}_{\text{atom}}}
\def\oHblack{\hat{\mathcal{H}}_{\text{black}}}
\def\oV{\hat{V}}
\def\oU{\hat{U}}
\def\oUd{\hat{U}^{\dagger}}
\def\oa{\hat{a}}
\def\oad{\hat{a}^{\dagger}}

\def\ophi{\hat{\phi}}
\def\oq{\hat{q}}
\def\opsi{\hat{\psi}}
\def\orho{\hat{\rho}}

\def\ovr{\hat{\bm{r}}}
\def\ovp{\hat{\bm{p}}}
\def\ovA{\hat{\bm{A}}}
\def\ovPi{\hat{\bm{\varPi}}}
\def\ovB{\hat{\bm{B}}}

\def\ovET{\hat{\bm{E}}_{\perp}}
\def\ovP{\hat{\bm{P}}}
\def\ovPT{\hat{\bm{P}}_{\perp}}
\def\ovDT{\hat{\bm{D}}_{\perp}}

\def\mdeltaT{\bm{\delta}_{\perp}}

\def\vr{\bm{r}}
\def\vR{\bm{R}}
\def\vA{\bm{A}}
\def\ve{\bm{e}}
\def\ovP{\bm{P}}
\def\vPT{\bm{P}_{\perp}}
\def\vDT{\bm{D}_{\perp}}

\def\vET{\bm{E}_{\perp}}

\def\ZZ{Z}
\def\ZZb{\bar{Z}}
\def\kB{k_{\text{B}}}

\def\psigmin{\psi_{\text{min},g}}
\def\psigdip{\psi_{\text{dip},g}}

\begin{document}

%\preprint{APS/123-QED}

\title{Circuit configurations which can/cannot
show super-radiant phase transitions}

\author{Motoaki Bamba}
\altaffiliation{E-mail: bamba@qi.mp.es.osaka-u.ac.jp}
\affiliation{Department of Materials Engineering Science, Osaka University, 1-3 Machikaneyama, Toyonaka, Osaka 560-8531, Japan}
\author{Nobuyuki Imoto}
\affiliation{Department of Materials Engineering Science, Osaka University, 1-3 Machikaneyama, Toyonaka, Osaka 560-8531, Japan}

\date{\today}

\begin{abstract}
Several superconducting circuit configurations are examined
on the existence of super-radiant phase transitions (SRPTs) in thermal equilibrium.
For some configurations
consisting of artificial atoms, whose circuit diagrams are however not specified,
and an LC resonator or a transmission line,
we confirm the absence of SRPTs in the thermal equilibrium
following the similar analysis as the no-go theorem for atomic systems.
We also show some other configurations
where the absence of SRPTs cannot be confirmed.
\end{abstract}

% PACS, the Physics and Astronomy Classification Scheme.

% 42.50.Ct	Quantum description of interaction of light and matter; related experiments
% 42.50.Pq	Cavity quantum electrodynamics; micromasers
% 05.30.Rt	Quantum phase transitions
% 64.70.Tg	Quantum phase transitions (for quantum Hall effects aspects, see 73.43.Nq in electronic structure of surfaces, interfaces, thin films, and low dimensional structures)
% 85.25.-j	Superconducting devices
% 42.50.-p	Quantum optics

%Use showkeys class option if keyword
                              %display desired
\maketitle
\section{Introduction}
A super-radiant phase transition (SRPT), i.e.,
a spontaneous appearance of (static) coherent amplitude of transverse electromagnetic fields
in the thermal equilibrium due to the light-matter interaction,
was first proposed theoretically around 1970 \cite{Mallory1969PR,Hepp1973AP,Wang1973PRA}.
It is different from the so-called super-radiance or super-fluorescence \cite{Gross1982PR},
i.e., a collective spontaneous emission from many atoms.
\ch{It is also different}
from the exciton super-radiance (one-photon super-radiance)
\cite{hanamura88}, i.e., an emission-rate enhancement by spatial broadening of wave-function of an excitation.
\ch{In contrast to these non-equilibrium phenomena,}
SRPTs are phase transitions in the thermal equilibrium.
Since the first proposals \cite{Mallory1969PR,Hepp1973AP,Wang1973PRA},
its absence (no-go theorem)
in atomic systems has been discussed based on the so-called $A^2$ term
\cite{Rzazewski1975PRL,Rzazewski1976PRA,Yamanoi1976PLA,Yamanoi1979JPA},
$P^2$ term \cite{Emeljanov1976PLA,Yamanoi1978P2},
gauge-invariance \cite{Woolley1976JPA,Knight1978PRA,Bialynicki-Birula1979PRA},
and minimal-coupling Hamiltonian \cite{Bialynicki-Birula1979PRA,Gawedzki1981PRA}.
Influences of the longitudinal dipole-dipole interaction
have also been discussed recently
\cite{Keeling2007JPCM,Vukics2014PRL,Bamba2014SPT,Vukics2015PRA,Griesser2016PRA}.

SRPTs require \ch{an} ultra-strong light-matter interaction \cite{Ciuti2005PRB,Devoret2007AP},
i.e., the interaction strength (vacuum Rabi splitting or absorption/emission rate in single-photon level)
must be comparable to or larger than frequencies
of electromagnetic waves and of transitions in matters.
In recent years, the ultra-strong interactions have been realized experimentally in a variety of systems
\cite{Gunter2009N,Anappara2009PRB,Todorov2009PRL,Todorov2010PRL,
Niemczyk2010NP,Fedorov2010PRL,Forn-Diaz2010PRL,
Schwartz2011PRL,Porer2012PRB,Scalari2012S,Goryachev2014PRA,Zhang2014PRLa,
Zhang2016NP,Forn-Diaz2017NP,Yoshihara2017NP}.
\ch{The presence of the so-called vacuum photons \cite{Ciuti2005PRB}
and the Schr\"odinger-cat-like state \cite{Ciuti2005PRB,Ashhab2010PRA}
are expected in the ground state under the ultra-strong interaction,
and recent experiments are indicating a signature of them
\cite{Yoshihara2017NP}.
However, the coherent amplitude of the electromagnetic fields
(expectation value of annihilation operator of a photon)
does not appear even in such a ground state,
but it is obtained only after a SRPT.
Currently,}
SRPTs are not yet observed experimentally in the thermal equilibrium,
while non-equilibrium analogues
were proposed theoretically \cite{Dimer2007PRA}
and observed experimentally in cold atoms driven by laser light
\cite{Baumann2010N,Baumann2011PRL}.

Instead of the atomic systems
\cite{Gunter2009N,Anappara2009PRB,Todorov2009PRL,Todorov2010PRL,
Schwartz2011PRL,Porer2012PRB,Scalari2012S,Zhang2016NP},
which are basically described by the minimal-coupling Hamiltonian
\cite{cohen-tannoudji89},
the possibility of the thermal-equilibrium SRPTs in superconducting circuits
\cite{Niemczyk2010NP,Fedorov2010PRL,Forn-Diaz2010PRL,Forn-Diaz2017NP,Yoshihara2017NP}
has been discussed \cite{Nataf2010NC,Viehmann2011PRL,Ciuti2012PRL,Jaako2016PRA,Bamba2016circuitSRPT}.
\ch{The existence of a SRPT was proposed for a superconducting circuit
with capacitive coupling between (two-level) artificial atoms and a resonator
by estimating the $A^2$ term to be relatively small \cite{Nataf2010NC}.
However, its estimation was doubted through
a standard description of superconducting circuit systems \cite{Viehmann2011PRL}.
After that, the existence of a SRPT was proposed again
for superconducting circuit with three-level artificial atoms
as a result of the modification of the sum rule
(and then of the $A^2$ term) \cite{Ciuti2012PRL}.
In these three works, their Hamiltonians were guessed
for standard circuit configurations
but without specifying circuit diagrams in detail,
although the derivation of exact Hamiltonians is crucial
for discussing the possibility of SRPTs.
Recently, the absence of SRPTs was confirmed
for a superconducting circuit diagram with capacitive coupling between
an LC resonator and charge qubits
by deriving its Hamiltonian in the standard quantization procedure
\cite{Jaako2016PRA}.
Almost at the same time, for a circuit diagram consisting of an LC resonator
coupled with Josephson junctions through inductors,
the existence of a SRPT was proposed
also in the standard quantization procedure \cite{Bamba2016circuitSRPT}.
No doubt is raised until now.
}

A remarkable feature of SRPTs is a decrease of the zero-point energy in the whole system
due to the light-matter interaction \cite{Emary2003PRL,Emary2003PRE,Nataf2010PRL,Bamba2016circuitSRPT}.
Chemical reactions \cite{Hutchison2012ACIE} and work functions \cite{Hutchison2013AM}
were reported to be modified by the ultra-strong interaction
with the vacuum electromagnetic fields.
The free energy, i.e., thermodynamic behaviors at finite temperatures,
should also be modified as suggested in Ref.~\cite{Canaguier-Durand2013ACIE},
while its experimental and theoretical evaluations are still under debate \cite{Sen2015PRB,Cwik2016PRA}.
In the superconducting circuit proposed in Ref.~\cite{Bamba2016circuitSRPT},
an external magnetic flux bias
or $\pi$ junctions \cite{Ryazanov2001PRL}
are inevitable for realizing the SRPT in the thermal-equilibrium.
The external magnetic flux increases the zero-point energy of the circuit.
While the zero-point energy is certainly decreased by the increase in the photon-atom interaction strength,
it cannot be lower than the \ch{zero-point energy} in the absence of the external magnetic flux.
It is still open to dispute whether there is a lower bound of the zero-point energy in superconducting circuits.
If there exists a superconducting circuit showing a SRPT without the external magnetic flux
or $\pi$ junctions,
the zero-point energy should be purely decreased by increasing
the strength of the interaction with the transverse electromagnetic fields,
and the thermodynamic properties,
e.g., the superconducting transition temperature,
of the circuit might be modified.

In order to find such a circuit structure, in this paper,
we show some hopeless circuit configurations
where SRPTs are absent even in the presence of an external magnetic flux or $\pi$ junctions.
There is a large number of degrees of freedom in designing circuit structures,
and there is not a standard Hamiltonian
corresponding to the minimal-coupling one for the atomic systems.
In order to rule out a wide range of circuit structures,
we treat artificial atoms as a black box,
i.e., we do not specify their circuit diagrams.
We consider some capacitive- and inductive-coupling configurations
between the black box and an LC resonator or a transmission line.
The absence of SRPTs in those configurations are confirmed
following the similar analysis as the no-go theorem for the atomic systems
\cite{Bialynicki-Birula1979PRA,Gawedzki1981PRA}
by deriving Hamiltonians in the flux- \cite{Devoret1997}
or charge-based \cite{Yurke1984PRA} standard quantization procedure.
\ch{In the analyses based on the $A^2$ term
\cite{Rzazewski1975PRL,Rzazewski1976PRA,Yamanoi1976PLA,Yamanoi1979JPA,Nataf2010NC,Viehmann2011PRL,Ciuti2012PRL},
on the $P^2$ term \cite{Emeljanov1976PLA,Yamanoi1978P2,Jaako2016PRA},
or on the softening of transition frequency \cite{Emary2003PRL,Emary2003PRE,Nataf2010NC,Bamba2016circuitSRPT},
we must specify circuit diagrams of whole systems in detail.
In contrast, in this paper, the artificial atoms are treated as a black box
following the no-go theorem \cite{Bialynicki-Birula1979PRA,Gawedzki1981PRA},
but we need to specify only the connection between the black box
and a resonator.}

We also show some other circuit configurations
where the absence of SRPTs in the thermal equilibrium cannot be confirmed.
The circuit structure proposed in Ref.~\cite{Bamba2016circuitSRPT}
is certainly included in these configurations.
While our analysis does not depend on
whether an external magnetic flux or $\pi$ junctions are absent or not,
it does not \ch{rule out the possibility of SRPTs without
the external magnetic flux and $\pi$ junctions.}

This paper is organized as follows.
We first review the no-go theorem for atomic systems
in Sec.~\ref{sec:review}.
Following the similar analysis, in Sec.~\ref{sec:circuit},
we show the absence of SRPTs in three circuit configurations
by deriving Hamiltonians without specifying circuit diagrams of artificial atoms.
In Sec.~\ref{sec:go}, we show some other configurations
where the absence of SRPTs cannot be confirmed.
The discussion is summarized in Sec.~\ref{sec:summary}.

\section{No-go theorem for atomic systems} \label{sec:review}
In this section,
we review the no-go theorem of SRPTs in atomic systems
described by the minimal-coupling Hamiltonian.
It was mainly discussed in Refs.~\cite{Bialynicki-Birula1979PRA,Gawedzki1981PRA}
based on the c-number substitution \cite{Wang1973PRA,Hepp1973PRA,Hemmen1980PLA,Gawedzki1981PRA},
which is also used in the semi-classical analysis of Ref.~\cite{Bamba2016circuitSRPT}.

The minimal-coupling Hamiltonian is expressed as \cite{cohen-tannoudji89}
\begin{align} \label{eq:oHmin_EB} % !!!!!!!!!!!!!!!!!!!!!!!!!!!!!!!!!!!!!!!!!!
\oHmin
& = \int\dd\vr \left\{ \frac{\diez\ovET(\vr)^2}{2} + \frac{\ovB(\vr)^2}{2\muz} \right\}
\nonumber \\ & \quad
+ \sum_{j=1}^N\frac{[\ovp_j-e_j\ovA(\ovr_j)]^2}{2m_j}
 + \oV(\{\ovr_j\}).
\end{align}
Here, the second last term is the kinetic energy
of charged particles. $N$ is the number of the particles.
$\ovr_j$ and $\ovp_j$ are operators of a position and a momentum,
respectively, of the $j$-th particle with a mass $m_j$ and a charge $e_j$.
They satisfy $[\ovr_j, \ovp_{j'}] = \delta_{j,j'}\ii\hbar\bm{1}$.
The last term $\oV$ represents the Coulomb interaction between the charged particles,
and it depends only on the particles' positions $\{\ovr_j\}$.
The first and second terms represent the energies
of the transverse electric field $\ovET(\vr)= - \ovPi(\vr)/\diez$
and the magnetic flux density $\ovB(\vr)=\rot\ovA(\vr)$, respectively.
Here, $\ovA(\vr)$ is the vector potential and $\ovPi(\vr)$ is its conjugate momentum satisfying
\begin{equation}
\left[ \ovA(\vr), \ovPi(\vr') \right] = \ii\hbar\mdeltaT(\vr-\vr'),
\end{equation}
where $\mdeltaT(\vr-\vr')$ is the transverse delta function \cite{cohen-tannoudji89}.
We rewrite these fields by annihilation and creation operators as
\begin{subequations}
\begin{align}
\ovA(\vr)
& = \sum_{k=1}^M \ve_k \sqrt{\frac{\hbar}{2\diez\omega_k}}f_k(\vr)
    \left( \oa_k + \oad_k \right), \label{eq:oA} \\ % !!!!!!!!!!!!!!!!!!!!!!!!
\ovPi(\vr)
& = - \sum_{k=1}^M \ve_k \ii \sqrt{\frac{\hbar\diez\omega_k}{2}}f_k(\vr)
    \left( \oa_k - \oad_k \right).
\end{align}
\end{subequations}
Here, $\oa_k$ annihilates a photon in the $k$-th
mode of the electromagnetic wave with a frequency of $\omega_k$.
$f_k(\vr)$ is the wavefunction of the $k$-th mode,
$\ve_k$ is the unit vector in its polarization direction,
and $\diez$ is the vacuum permittivity.
$M$ is the number of modes.
The minimal-coupling Hamiltonian in Eq.~\eqref{eq:oHmin_EB} is rewritten as
\begin{align}
\oHmin
& = \sum_{k=1}^M \hbar\omega_k \left(\oad_k\oa_k+\frac{1}{2}\right)
\nonumber \\ & \quad
+ \sum_{j=1}^N\frac{[\ovp_j-e_j\ovA(\ovr_j)]^2}{2m_j}
+ \oV(\{\ovr_j\}).
\label{eq:oHmin} % !!!!!!!!!!!!!!!!!!!!!!!!!!!!!!!!!!!!!!!!!!
\end{align}
For simplicity, as discussed in Ref.~\cite{Bialynicki-Birula1979PRA},
we apply the long-wavelength approximation (electric-dipole approximation),
i.e., the vector potential is rewritten as
\begin{equation}
\ovA(\ovr_j) \simeq \ovA(\vR_j),
\end{equation}
where $\vR_j$ is the rough position
of the $j$-th particle (e.g., position of lattice site).
The long-wavelength approximation is justified
when the amplitude of the vector potential
varies only slightly by the distance $\ovr_j - \vR_j$.
In other words, $\ovr_j - \vR_j$
is much shorter than the wavelength of the electromagnetic wave in the frequency range of interest.
A more general discussion beyond the long-wavelength approximation
is shown in Ref.~\cite{Gawedzki1981PRA}.

\ch{Expanding the kinetic energy of the charged particles
in Eq.~\eqref{eq:oHmin_EB} or Eq.~\eqref{eq:oHmin},
we get $- \sum_{j=1}^N(e_j/m_j)\ovp_j\cdot\ovA(\ovr_j)$
and $\sum_{j=1}^Ne^2\ovA(\ovr_j)^2/(2m_j)$.
The former leads to the light-matter interaction term,
and the latter leads to the $A^2$ term
\cite{Rzazewski1975PRL,Rzazewski1976PRA,Yamanoi1976PLA,Yamanoi1979JPA}.
The absence of SRPTs by the presence of the $A^2$ term can be confirmed
when we specify the atomic systems of interest,
especially the shape of $\oV(\{\ovr_j\})$.
In contrast, the following no-go theorem shows the absence of SRPTs
generally in the minimal-coupling Hamiltonian, i.e.,
without specifying the systems in detail.
}

The thermodynamic properties at a finite temperature $T$
is analyzed by the partition function for $\beta = 1/(\kB T)$ as
\begin{equation} \label{eq:Z_min_exact} % !!!!!!!!!!!!!!!!!!!!!!!!!!!!!!!!!!!
\ZZ(T) = \Tr\left[\ee^{-\beta\oHmin}\right].
\end{equation}
As discussed in Refs.~\cite{Wang1973PRA,Hepp1973PRA,Hemmen1980PLA,Gawedzki1981PRA},
we replace the trace over the photonic variables
by the integral over the coherent state as
\begin{equation} \label{eq:Z_minimal} % !!!!!!!!!!!!!!!!!!!!!!!!!!!!!!!!!!!!!!
\ZZb(T)
= \int\left(\prod_k\frac{\dd^2\alpha_k}{\pi}\right)
  \Tr\left[ \ee^{-\beta\oHmin'} \right],
\end{equation}
where the photon operators $\{\oa_k,\oad_k\}$ and vector potential $\ovA(\vR_j)$ are
replaced by c-numbers as
\begin{align} \label{eq:oHmin_c} % !!!!!!!!!!!!!!!!!!!!!!!!!!!!!!!!!!!!!!!!
\oHmin'
& = \sum_{k=1}^M \hbar\omega_k\left(|\alpha_k|^2+\frac{1}{2}\right)
\nonumber \\ & \quad
+ \sum_{j=1}^N\frac{[\ovp_j-e_j\vA(\vR_j)]^2}{2m_j} + \oV(\{\ovr_j\}).
\end{align}
Here, $\alpha_k\in\mathbb{C}$ is an amplitude of a coherent state $\ket{\alpha_k}_k$
in the $k$-th mode giving $\oa_k\ket{\alpha_k}_k = \alpha_k\ket{\alpha_k}_k$.
The c-number vector potential is expressed as
\begin{equation}
\vA(\vr) = \sum_{k=1}^M \ve_k \sqrt{\frac{\hbar}{2\diez\omega_k}}f_k(\vr)\left(\alpha_k+\alpha_k^*\right).
\end{equation}
The replacement (approximation) performed in Eq.~\eqref{eq:Z_minimal} is called the c-number substitution
\cite{Hemmen1980PLA,Gawedzki1981PRA},
and the analysis based on it is called the semi-classical analysis in Ref.~\cite{Bamba2016circuitSRPT},
since the photonic operators are treated as the c-numbers.

For justifying this c-number substitution,
we must consider the thermodynamic limit $N\to\infty$.
Further, in the early study by Wang and Hioe \cite{Wang1973PRA},
they note that this substitution is justified on the following two assumptions:
\begin{description}
\item[\it Assumption 1] The limits as $N\rightarrow\infty$ of the field operator $\oa/\sqrt{N}$
and $\oad/\sqrt{N}$ exist.
\item[\it Assumption 2] The order of the double limit in the exponential series
$\lim_{N\rightarrow\infty}\lim_{R\rightarrow\infty}\sum_{r=1}^R (-\beta \oH)^r/r!$
can be interchanged.
\end{description}
The first assumption implies that 
$\alpha_k/\sqrt{N}$ should be of a finite value after the SRPTs
in the thermodynamic limit $N\to\infty$.
On the other hand, it is hard to check the second assumption for arbitrary systems.
Instead, we follow the justification discussed in Ref.~\cite{Hepp1973PRA}.
The exact partition function $\ZZ(T)$ in Eq.~\eqref{eq:Z_min_exact}
and the approximated one $\ZZb(T)$ in Eq.~\eqref{eq:Z_minimal}
satisfy the following relation \cite{Hepp1973PRA}:
\begin{equation}
\ZZb(T) \leq \ZZ(T) \leq \exp\left(\frac{1}{\kB T}\sum_{k=1}^M\hbar\omega_k\right)\ZZb(T).
\end{equation}
From this, the free energy $-(\kB T/N)\ln\ZZ(T)$ per atom satisfies
\begin{multline}
-\frac{1}{N} \sum_{k=1}^M\hbar\omega_k
-\frac{\kB T}{N}\ln\ZZb(T)\\
\leq -\frac{\kB T}{N}\ln\ZZ(T)
\leq  -\frac{\kB T}{N}\ln\ZZb(T).
\end{multline}
Therefore, in the thermodynamic limit $N\to\infty$,
$\ZZ(T)$ is well approximated by $\ZZb(T)$,
if systems of interest satisfy
\begin{description}
\item[\it Assumption A] $\displaystyle\lim_{N\to\infty}\sum_{k=1}^M\frac{\hbar\omega_k}{N} \ll \left|\frac{\kB T}{N}\ln\ZZb(T)\right|$.
\end{description}
\ch{This condition can be checked when we specify atomic systems of interest.
It is satisfied for ensemble of two-level atoms
\cite{Hepp1973PRA}, i.e., in the Dicke Hamiltonian.
For superconducting circuits, it was checked numerically
for the circuit proposed in Ref.~\cite{Bamba2016circuitSRPT}.}
In this paper, we implicitly consider that the systems of interest satisfy
{\it Assumptions 1 and 2, or A} in the thermodynamic limit $N\to\infty$,
\ch{while we do not specify the systems in detail.}
In other words, we cannot discuss the absence of SRPTs
in systems that do not satisfy these assumptions,
since we cannot rewrite the partition function as Eq.~\eqref{eq:Z_minimal}
and the following analysis is not justified.

The no-go theorem \cite{Bialynicki-Birula1979PRA}
for atomic systems in the long-wavelength approximation
is discussed based on the partition function in Eq.~\eqref{eq:Z_minimal}
described by the minimal-coupling Hamiltonian in Eq.~\eqref{eq:oHmin_c}
under the c-number substitution.
If there exists a state $\ket{\psi(\{\alpha_k\})}$ that minimizes the energy
$\braket{\psi(\{\alpha_k\})|\oHmin'|\psi(\{\alpha_k\})}$
for a non-zero amplitude $\alpha_k \neq 0$,
the transverse electromagnetic fields get an amplitude spontaneously
in the ground state (and also in the thermal equilibrium for $T>0$),
i.e., the system shows a SRPT.
However, the absence of such a super-radiant ground state is confirmed
as seen in the following.

Here, we introduce a unitary operator
\begin{equation}
\oU_c \equiv \exp\left[ \frac{\ii}{\hbar} \sum_{j=1}^N e_j\ovr_j\cdot\vA(\vR_j) \right].
\end{equation}
Using this, we get
\begin{equation}
\oUd_c\ovp_j\oU_c = \ovp_j + e_j\vA(\vR_j).
\end{equation}
Then, since the Coulomb interaction $\oV$ does not depend on the momentum
$\{\ovp_j\}$ of the charged particles, we get
\begin{equation} \label{eq:UdHminU} % !!!!!!!!!!!!!!!!!!!!!!!!!!!!!!!!!!!!!!!!
\oHmin'' \equiv \oUd_c\oHmin'\oU_c
= \sum_{k=1}^M \hbar\omega_k\left( |\alpha_k|^2 + \frac{1}{2} \right)
+ \oHatom,
\end{equation}
where $\oHatom$ is the Hamiltonian of the charged particles
without the interaction with the transverse electromagnetic fields as
\begin{equation}
\oHatom
\equiv \sum_{j=1}^N\frac{\ovp_j{}^2}{2m_j} + \oV(\{\ovr_j\}).
\end{equation}
Since $\oU_c$ is a unitary operator,
the partition function in Eq.~\eqref{eq:Z_minimal} can be rewritten as
\begin{equation}
\ZZb(T)
= \int\left(\prod_k\frac{\dd^2\alpha_k}{\pi}\right)
  \Tr\left[ \ee^{-\beta\oHmin''} \right].
\end{equation}
Then, the problem is reduced to the minimization of
$\braket{\psi(\{\alpha_k\})|\oHmin''|\psi(\{\alpha_k\})}$
for trial state $\ket{\psi(\{\alpha_k\})}$.
Since $\oHatom$ in Eq.~\eqref{eq:UdHminU} is simply
the Hamiltonian of the charged particles,
the minimum energy is obtained for the following state:
\begin{equation}
\ket{\psigmin''} = \ket{\psi_g}_{\text{atom}}\otimes\ket{\{\alpha_k=0\}}_{\text{em}},
\end{equation}
where $\ket{\psi_g}_{\text{atom}}$ is the ground state of $\oHatom$
and $\ket{\{\alpha_k=0\}}_{\text{em}}$ represents a classical state
with zero amplitude for all the photonic modes.
In this way, the photonic modes do not spontaneously get an amplitude
in the ground state (and also in thermal equilibrium).
This is the basic logic of the no-go theorem of SRPTs in atomic systems
discussed in Refs.~\cite{Bialynicki-Birula1979PRA,Gawedzki1981PRA}.

On the other hand,
from the minimal-coupling Hamiltonian $\oHmin$ in Eq.~\eqref{eq:oHmin} without the c-number substitution,
we can get the Hamiltonian $\oHdipole$ of the length form \cite{cohen-tannoudji89,Keeling2007JPCM,Vukics2014PRL,Bamba2014SPT,Vukics2015PRA,Griesser2016PRA},
in contrast to $\oHmin$ called the velocity form.
Recovering the vector potential as an operator in the unitary operator as
\begin{equation}
\oU = \exp\left[ \frac{\ii}{\hbar} \sum_{j=1}^N e_j\ovr_j\cdot\ovA(\vR_j) \right],
\end{equation}
the Hamiltonian of the length form is obtained
in the long-wavelength approximation as
\begin{align}
\oHdipole & = \oUd\oHmin\oU \\
& = \int\dd\vr \left\{ \frac{[\ovDT(\vr)-\ovPT(\vr)]^2}{2\diez} + \frac{\ovB(\vr)^2}{2\muz} \right\}
  + \oHatom \label{eq:oHdip_2} \\ % !!!!!!!!!!!!!!!!!!!!!!!!!!!!!!!!!!!!!!!!!!
& = \sum_{k=1}^M \hbar\omega_k\left(\oad_k\oa_k+\frac{1}{2}\right)
- \frac{1}{\diez}\int\dd\vr\ \ovPT(\vr)\cdot\ovDT(\vr)
\nonumber \\ & \quad
+ \frac{1}{2\diez}\int\dd\vr\ \ovPT(\vr)^2
+ \oHatom. \label{eq:oHdip} % !!!!!!!!!!!!!!!!!!!!!!!!!!!!!!!!!!!!!!!!!!!!!!!
\end{align}
Here, $\ovPT(\vr)$ is the transverse component of the electric polarization
$\ovP(\vr) = \sum_j e_j\ovr_j\delta(\vr-\ovr_j)$,
while a more general definition is required beyond the long-wavelength approximation
(Power-Zienau-Woolley transformation) \cite{cohen-tannoudji89,Vukics2014PRL,Vukics2015PRA}.
The last term in the first line of Eq.~\eqref{eq:oHdip} represents the light-matter interaction
mediated by $\ovPT(\vr)$ and the transverse component of the electric displacement field
$\ovDT(\vr)$, which corresponds to the conjugate momentum of the vector potential as
$\ovDT(\vr) = - \ovPi(\vr)$ in the length form.
The second last term in Eq.~\eqref{eq:oHdip} is called the $P^2$ term,
by which the absence of SRPTs
\ch{can also been confirmed \cite{Emeljanov1976PLA,Yamanoi1978P2}
in the similar manner as the $A^2$ term.}

The ground state $\ket{\psigmin''}$ of $\oHmin''$
is not the exact ground state $\ket{\psigdip}$ of $\oHdipole$.
However, the absence of SRPTs itself can be confirmed as discussed above
if systems of interest satisfy {\it Assumptions 1 and 2, or A}
in the thermodynamic limit.
When the transverse electric polarization
$\vPT(\vr) = \braket{\psi_g|\ovPT(\vr)|\psi_g}_{\text{atom}}$
gets an amplitude
spontaneously in the ground state $\ket{\psi_g}_{\text{atom}}$ of the charged particles,
the electric displacement field can be induced as
$\vDT(\vr) = \vPT(\vr)$,
while the electric field is basically zero $\vET=(\vDT-\vPT)/\diez=0$,
by simply considering the minimization of the first term in Eq.~\eqref{eq:oHdip_2}
as the classical analysis in Ref.~\cite{Bamba2016circuitSRPT}.
Even though the photonic amplitude can get an amplitude as
$\braket{\psigdip|\ovPi(\vr)|\psigdip} \approx - \vDT(\vr)$
in the ground state of $\oHdipole$,
we do not call it a SRPT in this paper,
because the appearance of the photonic amplitude
originates from the system of charged particles $\oHatom$
not from the light-matter interaction.

% The above is the basic logic in the no-go theorem of SRPTs in atomic systems
% discussed in Refs.~\cite{Bialynicki-Birula1979PRA,Gawedzki1981PRA}.
While the possibility of SRPTs in atomic systems is still under debate
especially beyond the long-wavelength approximation
\cite{Gawedzki1981PRA,Vukics2015PRA,Griesser2016PRA},
the above logic is basically valid if the c-number substitution
performed in Eq.~\eqref{eq:Z_minimal} is justified,
i.e., if systems of interest satisfy {\it Assumptions 1 and 2, or A} in the thermodynamic limit $N\to\infty$.
Following this semi-classical analysis,
we examine the possibility of SRPTs in some superconducting circuit configurations
in the following sections.

\section{Circuit configurations where SRPTs are absent} \label{sec:circuit}
In this section,
we show three superconducting circuit configurations
where the absence of SRPTs can be confirmed
by the semi-classical analysis explained in the previous section.
Once we get an exact Hamiltonian of a circuit,
we can examine the possibility of SRPTs following the semi-classical analysis
or in other approaches \cite{Mallory1969PR,Hepp1973AP,Emary2003PRL,Emary2003PRE,Bamba2016circuitSRPT}.
However, in order to discuss a wide range of circuit structures,
Hamiltonians of general forms are preferred,
such as the minimal-coupling one for atomic systems.

\begin{figure}[tbp]
\includegraphics[scale=.33]{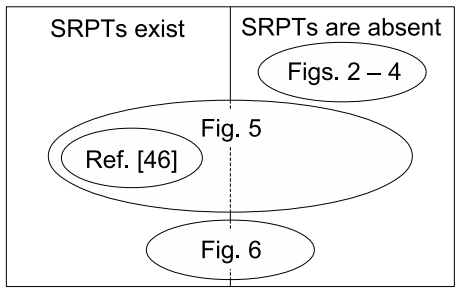}
\caption{Map of circuit configurations discussed in this paper
and the circuit proposed in Ref.~\cite{Bamba2016circuitSRPT}.
SRPTs are absent in the configurations depicted in Figs.~\ref{fig:2}--\ref{fig:4},
if systems of interest satisfy {\it Assumptions 1 and 2, or A}.
The absence of SRPTs cannot be confirmed in the configurations depicted in Figs.~\ref{fig:5} and \ref{fig:6}.
The circuit proposed in Ref.~\cite{Bamba2016circuitSRPT} \ch{[depicted in Fig.~\ref{fig:5}(c)]}
shows a SRPT and is included in the configurations of Figs.~\ref{fig:5}(a) and (b).
}
\label{fig:1}
\end{figure}
Figure \ref{fig:1} shows a map of circuit configurations
which we will discuss in this paper
and the circuit structure proposed in Ref.~\cite{Bamba2016circuitSRPT}.
We discuss the three circuit configurations
depicted in Figs.~\ref{fig:2}--\ref{fig:4}
with treating artificial atoms as a black box
(without specifying their circuit diagrams).
The absence of SRPTs will be confirmed
in an inductive-coupling configuration with an LC resonator in Sec.~\ref{sec:LC-L} (Fig.~\ref{fig:2}),
capacitive-coupling one with an LC resonator in Sec.~\ref{sec:LC-C} (Fig.~\ref{fig:3}),
and capacitive-coupling one with a transmission line in Sec.~\ref{sec:TL-C} (Fig.~\ref{fig:4}).
The two configurations depicted in Figs.~\ref{fig:5} and \ref{fig:6},
where the absence of SRPTs is not confirmed,
will be discussed in the next section.

\subsection{Inductive coupling with an LC resonator} \label{sec:LC-L}
\begin{figure}[tbp]
\includegraphics[scale=.33]{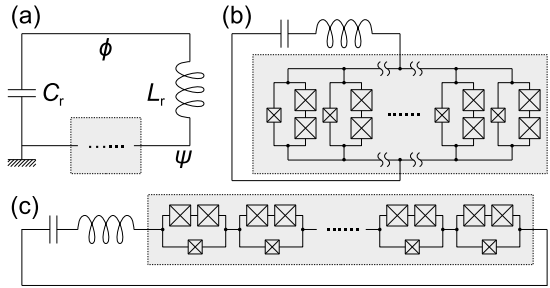}
\caption{(a) An LC resonator coupled inductively with a black box.
(b,c) Examples of circuits with artificial atoms.
This circuit configuration does not show SRPTs
by the coupling between the black box and the LC resonator.}
\label{fig:2}
\end{figure}
We first consider the circuit configuration depicted in Fig.~\ref{fig:2}(a)
consisting of a black box and an LC resonator with inductance $\LR$ and capacitance $\CR$.
Following the flux-based quantization procedure in Ref.~\cite{Devoret1997},
we define two node fluxes $\phi$, $\psi$ and the ground as Fig.~\ref{fig:2}(a).
A Lagrangian of this circuit is written as
\begin{equation} \label{eq:LL1} % !!!!!!!!!!!!!!!!!!!!!!!!!!!!!!!!!!!!!!!!!!!!
\LL_1 = \frac{\CR}{2}\dot{\phi}{}^2 - \frac{(\phi-\psi)^2}{2\LR}
  + \LLblack(\psi, \dot{\psi}; \ldots).
\end{equation}
The Lagrangian $\LLblack$ represents the elements in the black box,
and it is described by the flux $\psi$,
its time-derivative $\dot{\psi}$, and others inside the black box.
The conjugate momenta (charges) of $\phi$ and $\psi$ are derived, respectively, as
\begin{subequations}
\begin{align}
q & \equiv \frac{\partial\LL_1}{\partial \dot{\phi}} = \CR\dot{\phi}, \\
\rho
& \equiv \frac{\partial\LL_1}{\partial \dot{\psi}}
  = \frac{\partial\LLblack}{\partial \dot{\psi}}.
\end{align}
\end{subequations}
Then, we get a quantized Hamiltonian as
\begin{equation} \label{eq:oH1} % !!!!!!!!!!!!!!!!!!!!!!!!!!!!!!!!!!!!!!!!!!!!
\oH_1 = \frac{\oq^2}{2\CR} + \frac{(\ophi-\opsi)^2}{2\LR}
+ \oHblack(\opsi, \orho; \ldots),
\end{equation}
where $\oHblack$ is the Hamiltonian of the black box
derived from $\LLblack$.
The operators satisfy the following commutation relations:
\begin{subequations}
\begin{align}
[ \ophi, \oq ] & = \ii\hbar, \\
[ \opsi, \orho ] & = \ii\hbar,
\end{align}
\end{subequations}
and the other combinations are commutable.
We consider the flux $\ophi$ and the charge $\oq$
of the LC resonator as canonical variables of a photonic mode.
Introducing the annihilation operator $\oa$ of a photon
and an impedance $\ZR = \sqrt{\LR/\CR}$,
they are described as
\begin{subequations} \label{eq:ophi_oq_oa} % !!!!!!!!!!!!!!!!!!!!!!!!!!!!!!!!!
\begin{align}
\ophi & = \sqrt{\frac{\hbar\ZR}{2}}( \oa + \oad ), \\
\oq & = -\ii \sqrt{\frac{\hbar}{2\ZR}}( \oa - \oad ).
\end{align}
\end{subequations}
The resonance frequency is expressed as
\begin{equation}
\wc = \frac{1}{\sqrt{\LR\CR}}.
\end{equation}

In Eq.~\eqref{eq:oH1},
the coupling between the LC resonator and the black box
is described by the second term,
the inductive energy at $\LR$.
\ch{This expression corresponds to the Hamiltonian $\oHdipole$
of the length form in Eq.~\eqref{eq:oHdip_2}.
Expanding the second term,
we get $\ophi^2/(2\LR)$, $-\ophi\opsi/\LR$, and $\opsi^2/(2\LR)$
corresponding to the photonic flux energy,
the interaction term, and the $P^2$ term, respectively.
}

\ch{The no-go theorem for atomic systems starts
from the minimal-coupling Hamiltonian, Eq.~\eqref{eq:oHmin_EB},
where}
the light-matter coupling is described
by the kinetic term of the charged particles.
In order to describe the coupling by a part of the black box
as similar as the minimal-coupling Hamiltonian,
we transform Eq.~\eqref{eq:oH1} by a unitary operator
\begin{equation}
\oU_1 = \ee^{-\ii\oq\opsi/\hbar}.
\end{equation}
Using this operator, we get
\begin{subequations}
\begin{align}
\oUd_1\ophi\oU_1 & = \ophi + \opsi, \\
\oUd_1\orho\oU_1 & = \orho - \oq,
\end{align}
\end{subequations}
and the Hamiltonian is transformed to
\begin{subequations}
\begin{align}
\oH_{1'}
& \equiv \oUd_1\oH_1\oU_1 \\
& = \frac{\oq^2}{2\CR} + \frac{\ophi^2}{2\LR}
    + \oHblack(\opsi, \orho-\oq; \ldots) \\
& = \hbar\wc(\oad\oa + 1/2)
    + \oHblack(\opsi, \orho-\oq; \ldots).
\end{align}
\end{subequations}
This Hamiltonian has a similar form as the minimal-coupling Hamiltonian
in Eq.~\eqref{eq:oHmin}.
\ch{Specifying the black box and expanding the capacitive term depending on
$\orho-\oq$, such as $(\orho-\oq)^2/(2C)$ for a capacitance $C$,
we get an interaction term $-\orho\oq/C$ and the $A^2$ term $\oq^2/(2C)$.
However, the following discussion does not depend on the detail of the black box.}

Here, we suppose that there are many artificial atoms in the black box,
for example, as Figs.~\ref{fig:2}(b) and \ref{fig:2}(c),
and the circuit satisfies {\it Assumptions 1 and 2, or A}.
In the thermodynamic limit
(infinite number of artificial atoms; $N\rightarrow\infty$),
the partition function is written approximately as
\begin{equation}
\ZZb(T) = \int\frac{\dd^2\alpha}{\pi}\ \Tr\left[ \ee^{-\beta\oH'_{1'}} \right],
\end{equation}
where $\oa$ is replaced by a c-number $\alpha$ as
\begin{equation}
\oH'_{1'} = \hbar\wc(|\alpha|^2+1/2)
+ \oHblack(\opsi, \orho-q; \ldots),
\end{equation}
and the operator $\oq$ is also replaced by
\begin{equation}
q = -\ii \sqrt{\frac{\hbar}{2\ZR}}( \alpha - \alpha^* ).
\end{equation}
Here, by substituting the c-number also to the unitary operator as
\begin{equation}
\oU_{1c} = \ee^{-\ii q\opsi/\hbar},
\end{equation}
the partition function is rewritten as
\begin{equation}
\ZZb(T)
= \int\frac{\dd^2\alpha}{\pi}\ \Tr\left[ \ee^{-\beta\oH''_{1'}} \right],
\end{equation}
where
\begin{equation}
\oH''_{1'}
\equiv \oU_{1c} \oH'_{A'} \oUd_{1c}
= \hbar\wc(|\alpha|^2+1/2) + \oHblack(\opsi, \orho; \ldots).
\end{equation}
In this way, the problem is reduced to the similar one
discussed around Eq.~\eqref{eq:UdHminU} for atomic systems.
Then, SRPTs originating from the coupling
between the LC resonator and the black box \ch{are} absent
in the circuit configuration of Fig.~\ref{fig:2}(a),
if the circuits satisfy {\it Assumptions 1 and 2, or A}.

In Figs.~\ref{fig:2}(b) and \ref{fig:2}(c),
we suppose many flux qubits \cite{Orlando1999PRB},
which basically require an external magnetic flux in each loop
consisting of three Josephson junctions
for reaching the ideal two-level systems.
Even in the presence of the external magnetic fluxes in these loops,
the SRPTs are absent, because the Lagrangian is still expressed as Eq.~\eqref{eq:LL1},
while some phase transitions originating from the black box
(not from the coupling with LC resonator) can exist.
Of course, the SRPTs are absent
also when the external magnetic fluxes are completely absent.

\subsection{Capacitive coupling with an LC resonator} \label{sec:LC-C}
\begin{figure}[tbp]
\includegraphics[scale=.33]{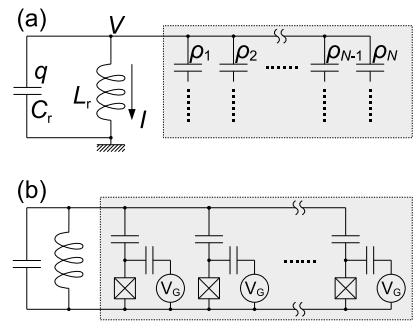}
\caption{(a) An LC resonator coupled capacitively with a black box.
(b) Example of circuits with artificial atoms,
which was already discussed in Ref.~\cite{Jaako2016PRA}.
This circuit configuration does not show SRPTs
by the coupling between the black box and the LC resonator.}
\label{fig:3}
\end{figure}
Next, we consider the circuit configuration depicted in Fig.~\ref{fig:3}(a).
An LC resonator couples with a black box through capacitances
inside the black box.
Following the charge-based quantization procedure in Ref.~\cite{Yurke1984PRA},
a Lagrangian is obtained as follows.
We define the ground, voltage $V$, current $I$, charges $q$ and $\{\rho_j\}$ for $j = 1, 2, \ldots, N$ as in Fig.~\ref{fig:3}(a).
The voltage $V$ and charge $q$ at capacitance $\CR$
are related as
\begin{equation}
V = \frac{q}{\CR}.
\end{equation}
The current $I$ through inductance $\LR$,
charges $\{\rho_j\}$ at coupling capacitances in the black box,
and $q$ at $\CR$ are related as
\begin{equation}
I = - \dot{q} - \sum_{j=1}^N\dot{\rho_j}.
\end{equation}
Further, the voltage $V$ and current $I$ are related as
\begin{equation}
V = \LR \dot{I}.
\end{equation}
Then, we get an equation of motion as
\begin{equation}
\ddot{q} + \sum_{j=1}^N\ddot{\rho_j} = \frac{q}{\LR\CR}.
\end{equation}
There are some other equations of motion describing the inside of the black box.
A Lagrangian giving these equations is in general represented as
\begin{equation}
\LL_2 = \frac{\LR}{2}\left(\dot{q}+\sum_{j=1}^N\dot{\rho}_j\right)^2 - \frac{q^2}{2\CR}
+ \LLblack(\{\rho_j\}, \{\dot{\rho}_j\}; \ldots).
\end{equation}
The conjugate momenta are derived as
\begin{subequations} \label{eq:conjugate2} % !!!!!!!!!!!!!!!!!!!!!!!!!!!!!!!!!!
\begin{align}
\phi & \equiv \frac{\partial\LL_2}{\partial\dot{q}}
= \LR\left(\dot{q}+\sum_{j=1}^N\dot{\rho}_j\right), \\
\psi_j & \equiv \frac{\partial\LL_2}{\partial\dot{\rho}_j}
= \LR\left(\dot{q}+\sum_{j=1}^N\dot{\rho}_j\right)
  + \frac{\partial\LLblack}{\partial\dot{\rho}_j}.
\end{align}
\end{subequations}
They satisfy
\begin{subequations}
\begin{align}
[ \oq, \ophi ] & = \ii\hbar, \\
[ \orho_j, \opsi_j ] & = \ii\hbar,
\end{align}
\end{subequations}
and other combinations are commutable.
The Hamiltonian is obtained as
\begin{equation} \label{eq:H2_drho} % !!!!!!!!!!!!!!!!!!!!!!!!!!!!!!!!!!!!!!!!
\HH_2 = \frac{\phi^2}{2\LR} + \frac{q^2}{2\CR}
+ \HHblack(\{\rho_j\}, \{\dot{\rho}_j\}; \ldots),
\end{equation}
where the Hamiltonian of the black box is defined as
\begin{align}
\HHblack(\{\rho_j\}, \{\dot{\rho}_j\}; \ldots)
& \equiv \sum_{j=1}^N\dot{\rho}_j\frac{\partial\LLblack}{\partial\dot{\rho}_j}
\nonumber \\ & \quad
  - \LLblack(\{\rho_j\},\{\dot{\rho}_j\}; \ldots).
\end{align}
Let us rewrite this in terms of $\{\rho_j\}$, $\{\psi_j\}$, \ldots.
From Eqs.~\eqref{eq:conjugate2}, we get
\begin{equation}
\frac{\partial\LLblack}{\partial\dot{\rho}_j}
= \psi_j - \phi.
\end{equation}
In the absence of the LC resonator,
we simply get $\partial\LLblack/\partial\dot{\rho}_j = \psi_j$,
and the Hamiltonian is represented as $\HHblack(\{\rho_j\}, \{\psi_j\}; \ldots)$.
Then, in the presence of the LC resonator,
$\psi_j$ is replaced by $\psi_j-\phi$ in $\HHblack$,
and the Hamiltonian in Eq.~\eqref{eq:H2_drho} is rewritten
in terms of $\{\rho_j\}, \{\psi_j\}, \ldots$ and in the quantized form as
\begin{equation}
\oH_2 = \frac{\ophi^2}{2\LR} + \frac{\oq^2}{2\CR}
+ \oH_{\text{black}}(\{\orho_j\}, \{\opsi_j-\ophi\}; \ldots).
\end{equation}
\ch{In this case, expanding an inductive energy depending on $\opsi_j-\ophi$
in the black box, we get an interaction term and the $A^2$ term,
when we specify the black box in detail.}
In the same manner as the previous subsection,
we rewrite $\ophi$ and $\oq$ by annihilation operator $\oa$
as
\begin{subequations} \label{eq:ophi_oq_oa_2} % !!!!!!!!!!!!!!!!!!!!!!!!!!!!!!!!!
\begin{align}
\ophi & = -\ii \sqrt{\frac{\hbar\ZR}{2}}( \oa - \oad ), \\
\oq & = \sqrt{\frac{\hbar}{2\ZR}}( \oa + \oad ).
\end{align}
\end{subequations}
Then, we replace $\oa$ by a c-number $\alpha$ as
\begin{equation}
\oH'_{2} = \hbar\wc(|\alpha|^2+1/2)
+ \oH_{\text{black}}(\orho, \opsi-\phi; \ldots),
\end{equation}
where $\ophi$ is also replaced by
\begin{equation}
\phi = - \ii \sqrt{\frac{\hbar\ZR}{2}}( \alpha - \alpha^* ).
\end{equation}
Using a unitary operator 
\begin{equation}
\oU_{2c} = \exp\left(\frac{\ii}{\hbar} \phi \sum_{j=1}^N\orho_j \right),
\end{equation}
we get
\begin{equation}
\oUd_{2c}\opsi_j\oU_{2c} = \opsi_j + \phi,
\end{equation}
and the problem is reduced to the minimization of 
\begin{align}
\oH''_{2}
& = \oUd_{2c} \oH'_{2} \oU_{2c}
\nonumber \\
& = \hbar\wc(|\alpha|^2+1/2) + \oH_{\text{black}}(\{\orho_j\}, \{\opsi_j\}; \ldots).
\end{align}
In the same manner as discussed above,
the SRPTs due to the coupling between the LC resonator and the black box
\ch{are} absent in the circuit configuration of Fig.~\ref{fig:3}(a),
if systems of interest satisfy {\it Assumptions 1 and 2, or A}.
Then, for example, the SRPTs are absent in the circuit of Fig.~\ref{fig:3}(b),
where the charge qubits couple capacitively with an LC resonator
as already discussed in Ref.~\cite{Jaako2016PRA}.

\subsection{Capacitive coupling with a transmission line} \label{sec:TL-C}
\begin{figure}[tbp]\begin{center}
\includegraphics[scale=.33]{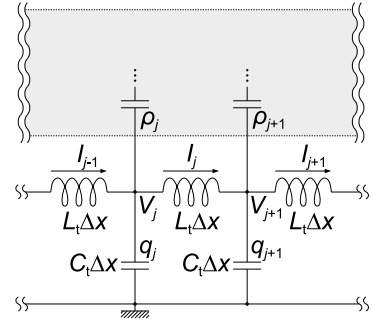}
\caption{A transmission line coupled capacitively with a long black box.
This circuit configuration does not show SRPTs
by the coupling between the black box and the transmission line.}
\label{fig:4}
\end{center}
\end{figure}
We next consider a transmission line
coupled capacitively with a long black box
as depicted in Fig.~\ref{fig:4}.
We can derive its Hamiltonian in the similar manner as the previous subsection.

In Fig.~\ref{fig:4},
$\CT$ and $\LT$ are, respectively, capacitance and inductance per unit length,
and $\dx$ is a short length for the discrete description of the transmission line.
We define voltage $V_j$, current $I_j$, and charges $q_j$ and $\rho_j$ as in Fig.~\ref{fig:4}.
The voltage $V_j$ and the charge $q_j$ at the $j$-th capacitance $\CT\dx$ is related as
\begin{equation}
V_j = \frac{q_j}{\CT\dx}.
\end{equation}
The current $I_j$ are related with the charges $q_j$ and $\rho_j$ as
\begin{equation}
I_j = I_{j-1} - \dot{q}_j - \dot{\rho}_j.
\end{equation}
Further, the voltage $V_j$ and the current $I_j$ are related as
\begin{equation}
V_{j+1} - V_{j} = - \LT\dx\dot{I}_j.
\end{equation}
From these relations, we get a difference equation as
\begin{equation}
\ddot{q}_j + \ddot{\rho}_j = \frac{q_{j+1} + q_{j-1} - 2q_j}{\LT\CT(\dx)^2}.
\end{equation}
This equation can be obtained by the following Lagrangian:
\begin{align}
\LL_3
& = \sum_j\left[ \LT\dx\left(\dot{q}_j+\dot{\rho}_j\right)^2
    - \frac{(q_{j+1}-q_j)^2}{2\CT\dx} \right]
\nonumber \\ & \quad
  + \LLblack(\{\rho_j\},\{\dot{\rho}_j\}; \ldots).
\end{align}
The conjugate momenta are derived as
\begin{subequations}
\begin{align}
\phi_j & \equiv \frac{\partial\LL_3}{\partial\dot{q}_j} = \LT\dx(\dot{q}_j + \dot{\rho}_j), \\
\psi_j & \equiv \frac{\partial\LL_3}{\partial\dot{\rho}_j} = \LT\dx(\dot{q}_j + \dot{\rho}_j)
+ \frac{\partial\LLblack}{\partial\dot{\rho}_j}.
\end{align}
\end{subequations}
Then, in the same manner as the previous subsection,
the Hamiltonian is derived as
\begin{align}
\HH_3
& = \sum_j\left[ \frac{\LT\dx}{2}\left(\dot{q}_j+\dot{\rho}_j\right)^2 + \frac{(q_{j+1}-q_j)^2}{2\CT\dx} \right]
\nonumber \\ & \quad
    + \HHblack(\{\rho_j\}, \{\dot{\rho}_j\}; \ldots), \\
\oH_3
& = \sum_j\left[ \frac{\ophi_j{}^2}{2\LT\dx} + \frac{(\oq_{j+1}-\oq_j)^2}{2\CT\dx} \right]
\nonumber \\ & \quad
    + \oHblack(\{\orho_j\}, \{\opsi_j-\ophi_j\}; \ldots). \label{eq:oH3} % !!
\end{align}
The first two terms are simply the Hamiltonian
of the transmission line,
in which a photon (microwave) propagates with a speed of $v = 1/\sqrt{\LT\CT}$
in the one-dimensional system.
The boundary conditions of the transmission line do not affect the possibility of SRPTs
in the semi-classical analysis relying on the c-number substitution.

In order to justify the c-number substitution performed in Eq.~\eqref{eq:Z_minimal},
let us discuss when the systems with the transmission line
satisfy {\it Assumption A}.
Here, we consider that the transmission line has a length of $\ell$.
The frequency of the photonic mode is $\omega_k = k(\pi v/\ell)$
for $k=1, 2, \ldots$.
Considering the minimum wavelength $\lambda_{\text{min}}$
where the electromagnetic wave interacts sufficiently with the artificial atoms
and is confined sufficiently in the one-dimensional transmission line,
the effective number of the photonic modes is determined as
$M=\ell/\lambda_{\text{min}}$.
The free energy per atom is in the same order as
the characteristic frequency $\wa$ of the atomic transition,
which gives a wavelength of $\lambda_{\text{a}} = 2\pi v/\wa$.
Instead of the limit $N\to\infty$,
we consider the limit of the number of atoms in the length of $\lambda_{\text{a}}$
as $n=N\lambda_{\text{a}}/\ell\to\infty$.
Then, {\it Assumption A} is rewritten as
\begin{align}
\frac{1}{N}\frac{\hbar\pi v}{\ell}\frac{M(M+1)}{2} & \ll \hbar\wa, \\
\frac{(\lambda_{\text{a}}/\lambda_{\text{min}}){}^2}{4} & \ll n.
\end{align}
In this way, the c-number substitution is justified
when the number $n$ of atoms in $\lambda_{\text{a}}$ is much lager than
$(\lambda_{\text{a}}/\lambda_{\text{min}})^2$.

In the same manner as the previous subsections,
when the c-number substitution is justified under the above condition,
the SRPTs due to the coupling between the transmission line
and the black box are absent in the circuit configuration of Fig.~\ref{fig:4}.

\section{Circuit configurations where SRPTs can exist} \label{sec:go}
Next, we show some circuit configurations where the absence of SRPTs
cannot be confirmed by the analysis in this paper.
In Sec.~\ref{sec:LC-L2} (Fig.~\ref{fig:5}), we discuss another inductive-coupling configuration
with an LC resonator.
In Sec.~\ref{sec:TL-L} (Fig.~\ref{fig:6}), an inductive-coupling configuration
with a transmission line is discussed.
As shown in Fig.~\ref{fig:1}, these configurations include
also the circuit structures that do not show SRPTs,
while the configuration of Fig.~\ref{fig:5}
includes the circuit proposed in Ref.~\cite{Bamba2016circuitSRPT}
that shows a SRPT.

\subsection{Another inductive coupling with an LC resonator} \label{sec:LC-L2}
\begin{figure}[tbp]
\begin{center}
\includegraphics[scale=.33]{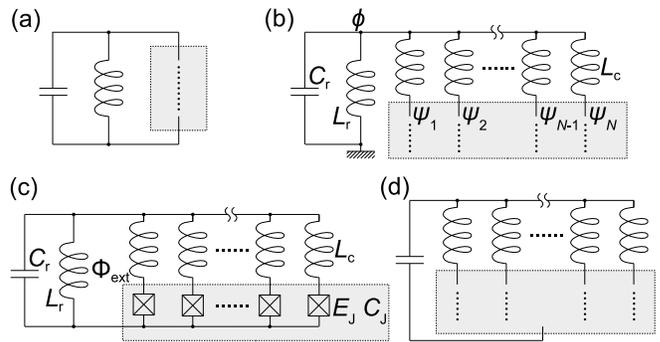}
\caption{An LC resonator coupled with a black box,
where the absence of SRPTs cannot be confirmed by the analysis in this paper.
It is because we could not derive a Hamiltonian for (a).
For (b) and \ch{(d)},
their Hamiltonians can be derived,
but they cannot be transformed as the minimal-coupling Hamiltonian.
\ch{(c) is the circuit proposed in Ref.~\cite{Bamba2016circuitSRPT}.}}
\label{fig:5}
\end{center}
\end{figure}
Let us first consider the circuit configuration depicted in Fig.~\ref{fig:5}(a),
which is generalized from the capacitive-coupling configuration in Fig.~\ref{fig:3}(a).
We could not derive a Hamiltonian of this configuration
in the flux- \cite{Devoret1997} or charge-based \cite{Yurke1984PRA} quantization procedure.
While other quantization procedures \cite{Solgun2014PRB} might give a Hamiltonian,
it in fact includes the circuit
\ch{of Fig.~\ref{fig:5}(c)}
proposed in Ref.~\cite{Bamba2016circuitSRPT},
which shows a SRPT in the presence of an external magnetic flux or $\pi$ junctions.
Then, even if we get a Hamiltonian of the circuit configuration in Fig.~\ref{fig:5}(a),
the absence of SRPTs would not be confirmed by the semi-classical analysis.

For example, let us consider the configuration in Fig.~\ref{fig:5}(b),
which is less general than Fig.~\ref{fig:5}(a)
but includes the circuit \ch{of Fig.~\ref{fig:5}(c) proposed} in Ref.~\cite{Bamba2016circuitSRPT}.
Following the flux-based procedure,
we define the ground and node fluxes $\phi$ and $\{\psi_j\}$ as in Fig.~\ref{fig:5}(b).
In the same manner as Sec.~\ref{sec:LC-L}, a Hamiltonian can be derived as
\begin{align} \label{eq:oH4} % !!!!!!!!!!!!!!!!!!!!!!!!!!!!!!!!!!!!!!!!!!!!!!!
\oH_4
& = \frac{\oq^2}{2\CR} + \frac{\ophi^2}{2\LR}
  + \sum_{j=1}^N\frac{(\ophi-\opsi_j)^2}{2\Lc}
\nonumber \\ & \quad
+ \oHblack(\{\opsi_j\}, \{\orho_j\}; \ldots).
\end{align}

\ch{
Let us derive the black-box Hamiltonian and roughly check the existence of the SRPT
for the circuit proposed in Ref.~\cite{Bamba2016circuitSRPT}
by specifying the detail inside the black box as Fig.~\ref{fig:5}(c).
Each $\Lc$ is connected with a Josephson junction
with Josephson energy $\EJ$ and capacitance $\CJ$.
A half of flux quantum $\fluxq = h/(2e)$ is applied to a loop
as an external flux bias $\Phiext = \fluxq/2$.
For this circuit, the black-box Hamiltonian is derived as \cite{Bamba2016circuitSRPT}
\begin{equation}
\oHblack^{\text{Ref.~\cite{Bamba2016circuitSRPT}}}(\{\opsi_j\}, \{\orho_j\})
= \sum_{j=1}^N\left( \frac{\orho_j{}^2}{2\CJ} + \EJ\cos\frac{2\pi\opsi_j}{\fluxq} \right).
\end{equation}
The sign of the last term (potential energy of the Josephson effect)
is positive by the presence of the external flux bias $\Phiext = \fluxq/2$.
We can intuitively understand the existence of a SRPT by analyzing the minima
of the inductive energy:
\begin{equation}
U(\phi, \psi)
= \frac{\phi^2}{2\LR} + \sum_{j=1}^N\left[
    \frac{(\phi-\psi_j)^2}{2\Lc} + \EJ\cos\frac{2\pi\psi_j}{\fluxq}
  \right].
\end{equation}
For $N\LR > [\fluxq/(2\pi)]^2/\EJ - \Lc$,
this function has two minima at $\phi = \pm \phi_0 \neq 0$
(and $\psi_j = \pm[1+\Lc/(N\LR)]\phi_0 \neq 0$).
Since the potential barrier between the two minima becomes infinitely high
in the thermodynamic limit $N\to\infty$,
the symmetry (superposition of the two minima) in the ground state
is broken spontaneously,
and we get a coherent amplitude of the flux $\phi \approx \pm \phi_0$
below a critical temperature.
In this way, SRPTs exist in superconducting circuits
where the photonic harmonic energy
[$\phi^2/(2\LR)$ minimized at $\phi = 0$]
and the atomic anharmonic energy
[$\EJ\cos(2\pi\psi_j/\fluxq)$ minimized at $\psi_j\neq0$]
competes through the coupling term
[$(\phi-\psi_j)^2/(2\Lc)$ minimized for $\phi = \psi_j$].
}

\ch{As we already found a counter-example above,
we cannot get the no-go theorem for the Hamiltonian
in Eq.~\eqref{eq:oH4} derived for the circuit in Fig.~\ref{fig:5}(b).}
In contrast to Sec.~\ref{sec:LC-L},
we cannot relocate the photonic flux $\phi$ into $\oHblack$
by unitary transformations,
since there are $N$ coupling terms $(\ophi-\opsi_j)^2/(2\Lc)$,
while the absence of SRPTs can be shown for $N=1$
in the same manner as Sec.~\ref{sec:LC-L}.
On the other hand, if we consider the third term, the inductive energies at $\Lc$,
as a part of the black-box Hamiltonian as
\begin{align}
& \oHblack'(\ophi;\{\opsi_j\}, \{\orho_j\}; \ldots)
\nonumber \\
& \equiv \sum_{j=1}^N\frac{(\ophi-\opsi_j)^2}{2\Lc}
+ \oHblack(\{\opsi_j\}, \{\orho_j\}; \ldots),
\end{align}
the coupling term is certainly included in the black box as
\begin{equation}
\oH_4
= \frac{\oq^2}{2\CR} + \frac{\ophi^2}{2\LR}
  + \oHblack'(\ophi;\{\opsi_j\}, \{\orho_j\}; \ldots).
\end{equation}
However, we cannot remove the photonic flux $\phi$
from the black-box Hamiltonian even under the c-number substitution.
For example, by introducing a unitary operator as
\begin{equation}
\oU_{4c} = \exp\left( -\frac{\ii}{\hbar}\phi\sum_{j=1}^N\orho_j\right),
\end{equation}
the Hamiltonian $\oH_4'$ under the c-number substitution is transformed to
\begin{equation}
\oUd_{4c}\oH'_4\oU_{4c}
= \frac{q^2}{2\CR} + \frac{\phi^2}{2\LR}
+ \oHblack''(\phi;\{\opsi_j\}, \{\orho_j\}; \ldots),
\end{equation}
where the black-box Hamiltonian is transformed as
\begin{align}
& \oHblack''(\phi;\{\opsi_j\}, \{\orho_j\}; \ldots)
\nonumber \\
& = \sum_{j=1}^N\frac{\opsi_j{}^2}{2\Lc}
+ \oHblack(\{\opsi_j+\phi\}, \{\orho_j\}; \ldots).
\end{align}
In this way, the problem cannot be reduced to the minimization
of the black-box Hamiltonian without the LC resonator.
In other words, the Hamiltonian of the circuit configuration in Fig.~\ref{fig:5}(b)
cannot be expressed as similar as the minimal-coupling Hamiltonian.
Then, the absence of SRPTs cannot be confirmed by the same logic as the no-go theorem
for atomic systems.
This result is consistent with the proposal of a SRPT in Ref.~\cite{Bamba2016circuitSRPT}.

In the similar manner,
for the circuit configuration of Fig.~\ref{fig:5}\ch{(d)},
where $\LR$ is eliminated,
its Hamiltonian is simply derived as Eq.~\eqref{eq:oH4}
without the second term.
The absence of SRPTs cannot be confirmed
also in this circuit configuration.

\subsection{Inductive coupling with a transmission line} \label{sec:TL-L}
\begin{figure}[tbp]\begin{center}
\includegraphics[scale=.33]{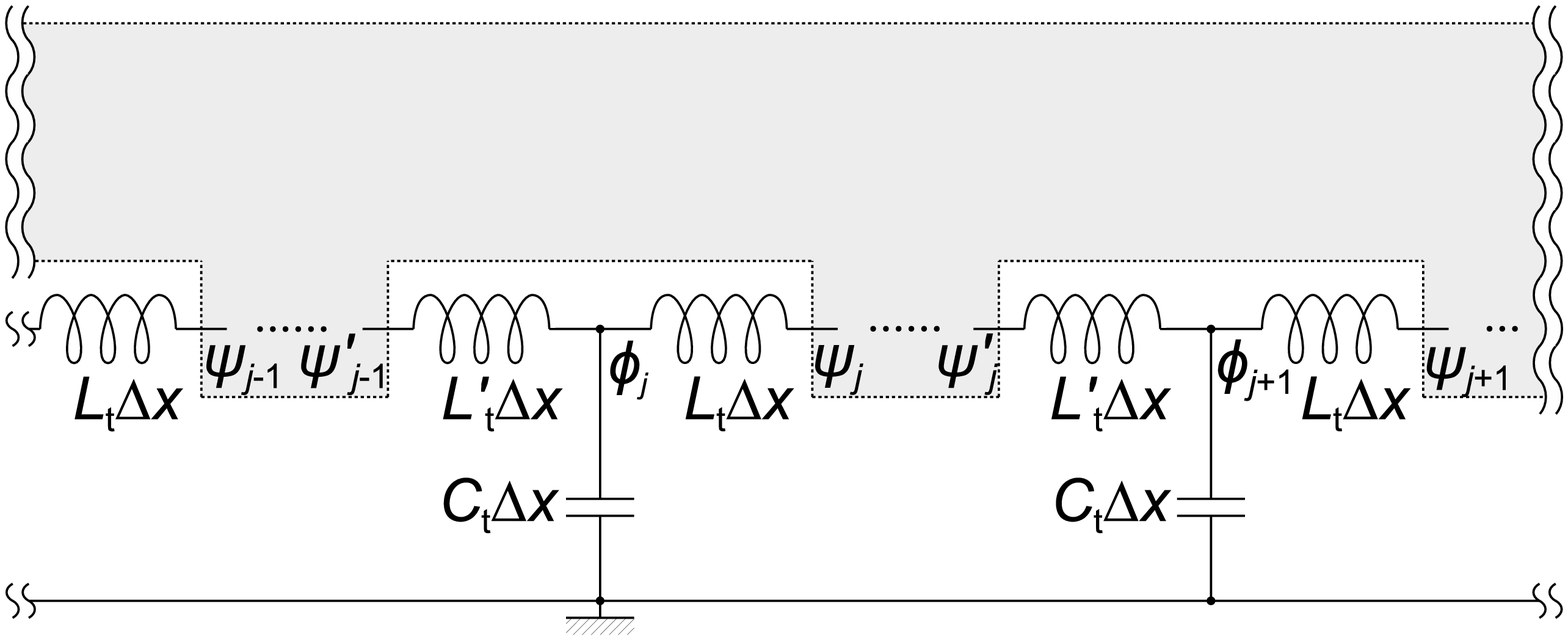}
\caption{A transmission line coupled inductively with a long black box.
The absence of SRPTs cannot be confirmed by the analysis in this paper.}
\label{fig:6}
\end{center}
\end{figure}
Finally, let us consider the circuit configuration depicted in Fig.~\ref{fig:6}.
A transmission line couples with a long black box inductively,
or we can instead consider small LC resonators coupled through the black box.
Following the flux-based procedure, a Lagrangian is obtained as
\begin{align}
\LL_5
& = \sum_j\left[ \frac{\CT\dx}{2}\dot{\phi}_j{}^2
- \frac{(\phi_j-\psi_j)^2}{2\LT\dx}
- \frac{(\phi_j-\psi'_{j-1})^2}{2\LT'\dx} \right]
\nonumber \\ & \quad
+ \LLblack(\{\psi_j\}, \{\dot{\psi}_j\}; \{\psi'_j\}, \{\dot{\psi}'_j\}; \ldots).
\end{align}
The conjugate momenta are derived as
\begin{subequations}
\begin{align}
q_j & \equiv \frac{\partial \LL_5}{\partial\dot{\phi}_j} = \CT\dx\dot{\phi}_j, \\
\rho_j & \equiv \frac{\partial \LL_5}{\partial\dot{\psi}_j}
  = \frac{\partial \LLblack}{\partial\dot{\psi}_j}, \\
\rho'_j & \equiv \frac{\partial \LL_5}{\partial\dot{\psi}'_j}
  = \frac{\partial \LLblack}{\partial\dot{\psi}'_j}.
\end{align}
\end{subequations}
Then, we get the Hamiltonian as
\begin{align}
\oH_5
& = \sum_j\left[ \frac{\oq_j{}^2}{2\CT\dx} + \frac{(\ophi_j-\opsi_j)^2}{2\LT\dx}
    + \frac{(\ophi_j-\opsi'_{j-1})^2}{2\LT'\dx} \right]
\nonumber \\ & \quad
  + \oHblack(\{\opsi_j\}, \{\orho_j\}; \{\opsi'_j\}, \{\orho'_j\}; \ldots).
\end{align}
For this Hamiltonian, we cannot relocate the coupling terms
into $\oHblack$ as in the previous sections.
For example, using a unitary operator
\begin{equation}
\oU_5 = \exp\left(- \frac{\ii}{\hbar}\sum_j\oq_j\opsi_j\right),
\end{equation}
we get
\begin{subequations}
\begin{align}
\oUd_5\ophi_j\oU_5 & = \ophi_j + \opsi_j, \\
\oUd_5\orho_j\oU_5 & = \orho_j - \oq_j,
\end{align}
\end{subequations}
and
\begin{align}
& \oUd_5\oH_5\oU_5 \nonumber \\
& = \sum_j\left[ \frac{\oq_j{}^2}{2\CT\dx} + \frac{\ophi_j{}^2}{2\LT\dx}
    + \frac{(\ophi_j+\opsi_j-\opsi'_{j-1})^2}{2\LT'\dx} \right]
\nonumber \\ & \quad
  + \oHblack(\{\opsi_j\}, \{\orho_j-\oq_j\}; \{\opsi'_j\}, \{\orho'_j\}; \ldots). \label{eq:UdHU4}
\end{align}
In this way, the coupling terms inevitably remains in the photonic Hamiltonian
as far as we tried.
Then, the absence of SRPTs in the transmission line
of Fig.~\ref{fig:6} cannot be confirmed by the analysis in this paper,
while its Hamiltonian could be derived with treating the artificial atoms as a black box.

\section{Summary} \label{sec:summary}
Following the similar analysis as the no-go theorem for atomic systems
\cite{Bialynicki-Birula1979PRA,Gawedzki1981PRA},
we examined the possibility of SRPTs in some configurations of superconducting circuits.
By deriving Hamiltonians with treating artificial atoms as a black box,
we show that three configurations depicted in Figs.~\ref{fig:2}--\ref{fig:4}
do not show SRPTs if the systems satisfy {\it Assumptions 1 and 2, or A} in the thermodynamic limit,
which justify the c-number substitution performed in Eq.~\eqref{eq:Z_minimal}
and are essential in the no-go theorem for the atomic systems \cite{Bialynicki-Birula1979PRA,Gawedzki1981PRA}.

The absence of SRPTs cannot be confirmed for the circuit configurations
in Figs.~\ref{fig:5} and \ref{fig:6}.
It is because, for Fig.~\ref{fig:5}(a), we could not derive its Hamiltonian
with treating artificial atoms as a black box.
Concerning Figs.~\ref{fig:5}(b), \ch{(d)}, and \ref{fig:6},
we can derive their Hamiltonians,
but they cannot be transformed as the minimal-coupling Hamiltonian.
Then, the absence of SRPTs cannot be confirmed in the analysis of this paper.
In fact, Figs.~\ref{fig:5}(a) and (b) includes the circuit
in Ref.~\cite{Bamba2016circuitSRPT} \ch{depicted in Fig.~\ref{fig:5}(c)},
where a SRPT in the thermal equilibrium was proposed
in the presence of an external magnetic flux or $\pi$ junctions.

The analysis in this paper shows the absence of SRPTs
originating from the coupling between the black box
and the LC resonator or the transmission line.
If the black box includes another resonator or transmission line,
we must examine whether it can be reduced to the three circuit configurations
in Fig.~\ref{fig:2}--\ref{fig:4}
or we must extend the discussion for circuits
with multiple resonators or transmission lines.
Further, there also remains the possibility of SRPTs
in systems that do not satisfy {\it Assumption 1, 2, or A},
i.e., those SRPTs cannot be analyzed under the c-number substitution
performed in Eq.~\eqref{eq:Z_minimal}.

In order to find SRPTs in the absence of an external magnetic flux or $\pi$ junctions,
we should explore the circuit configurations in Figs.~\ref{fig:5} and \ref{fig:6}
or others except Figs.~\ref{fig:2}--\ref{fig:4},
while the analysis in this paper does not basically depend on
whether an external magnetic flux or $\pi$ junctions exist or not.

\begin{acknowledgments}
M.~B.~ thanks P.-M.~Billangeon for fruitful discussions.
This work was funded by ImPACT Program of Council for Science, Technology and
Innovation (Cabinet Office, Government of Japan)
and by KAKENHI (Grants No.~26287087, No.~JP16H02214, and No.~24-632).
\end{acknowledgments}

% \bibliographystyle{bamba_prl}
% \bibliography{../../../../OneDrive/bib/list,../../../../OneDrive/bib/bamba}

\end{document}